\documentstyle[seceq,preprint]{ptptex}
\def\tr{\mathop{\rm tr}\nolimits}

\makeatletter
\@addtoreset{equation}{section}
\makeatother

\newcommand{\VEV}[1]{\left\langle #1 \right\rangle}

\newcommand{\nn}{\nonumber}

\newcommand{\order}[1]{{\cal O}(#1)}

\newcommand{\GeV}{\mbox{GeV}}

\newcommand{\ie}{{\it i.e.}}
\newcommand{\eg}{{\it e.g.}}

\newcommand{\E}[1]{$E_#1$}

\newcommand{\abs}[1]{\left| #1 \right|}

\newcommand{\cc}[1]{\overline{#1}}

\newcommand{\bequ}{\begin{equation}}
\newcommand{\eequ}{\end{equation}}
\newcommand{\beqn}{\begin{eqnarray}}
\newcommand{\eeqn}{\end{eqnarray}}
\newcommand{\bctr}{\begin{center}}
\newcommand{\ectr}{\end{center}}

\newcommand{\Ll}{\left[}
\newcommand{\Rl}{\right]}

\newcommand{\II}{I$\!$I}
\newcommand{\III}{I$\!$I$\!$I}
\newcommand{\IV}{I$\!$V}
\notypesetlogo  %comment in if to eliminate PTPTeX logo
\preprintnumber[3cm]{%<-- [..]: optional width of preprint # column.
KUNS-1830\\ hep-ph/0303207}
\title{%        %You can use \\ for explicit line-break
Simple $E_6$ Unification with Anomalous $U(1)_A$ Symmetry
}
\author{%       %Use \sc for the family name
Nobuhiro {\sc Maekawa}\footnote{
E-mail: maekawa@gauge.scphys.kyoto-u.ac.jp}
and 
Toshifumi {\sc Yamashita}\footnote{
E-mail: yamasita@gauge.scphys.kyoto-u.ac.jp}
}

\inst{%         %Affiliation, neglected when [addenda] or [errata]
Department of Physics, Kyoto University, Kyoto 606-8502, Japan
}

\recdate{%      %Editorial Office will fill in this.
}

\abst{
We propose simpler Higgs sectors in $E_6$ grand unified theory%
~(GUT) with anomalous $U(1)_A$ gauge symmetry than the previous
model in Ref.\citen{E6}. 
As in the previous model,
the doublet-triplet~(DT) splitting can be realized in a natural way, 
while proton decay via dimension 5 operators is suppressed and 
gauge coupling unification is also realized without fine-tuning.
Combining the matter sector, simple complete GUTs can be obtained.
Since the Higgs sector is simpler, the gauge coupling constant
at the cutoff scale can be in perturbative region, and therefore,
the estimated value of the lifetime of a nucleon in this model, 
$\tau_p(p\rightarrow e^+\pi^0)\sim 5\times 10^{33}$ years, becomes
more reliable.
}

\begin{document}

\maketitle

%--------------------<<   section    >>--------------------
\section{Introduction}
In a series of papers,\cite{E6,NGCU,maekawa2,horiz,BM,maekawa} 
it has been understood that anomalous $U(1)_A$ gauge symmetry,%
\cite{U(1)} whose anomaly is cancelled by the Green-Schwarz 
mechanism,\cite{GS} plays an important role in  solving various 
problems in GUTs. One of the most important features of
this scenario is to introduce generic interactions even for higher 
dimensional operators with $\order1$ coefficients.
Of course, generic interactions are often introduced in matter
sector,\cite{Matter} but not in Higgs sector. The generic interactions
even in the Higgs sector are the 
biggest difference between our scenario and the previous GUT scenarios
with (anomalous) $U(1)$ symmetry.\cite{Higgs}
Therefore, once we fix the symmetry of the theory, we can define 
the theory except for the $\order1$ coefficients. 
Because the gauge symmetry is $SO(10)\times U(1)_A$ or 
$E_6\times U(1)_A$, the parameters to fix the 
symmetry are essentially the (integer) charges of the $U(1)_A$ 
for the fields introduced in the  theory.
It is surprising that only by setting the symmetry, we can obtain 
complete GUTs, in which the DT splitting\cite{DTsplitting} 
is realized without too rapid proton decay via dimension 5 operators, 
the realistic structure of quark and lepton 
mass matrices is obtained including bi-large neutrino mixings\cite{SK}
by using the Froggatt-Nielsen~(FN) mechanism,\cite{FN}  
and the natural gauge coupling unification is realized.\cite{NGCU}
As a result of the natural gauge coupling unification, 
the cutoff scale $\Lambda$ is around 
the usual GUT scale $\Lambda_G\sim 2\times 10^{16}$ GeV and 
the gauge couplings are unified just below $\Lambda$.
Therefore more rapid proton decay via dimension 6 operators,
$p\rightarrow e^+\pi^0$, is one of the most interesting prediction
in the scenario.
And if we include SUSY breaking sector, the $\mu$ problem can also 
naturally be solved.\cite{maekawa2} 
Moreover, if we introduce a horizontal non-abelian gauge symmetry,
$SU(2)_H$ or $SU(3)_H$, then the SUSY flavor problem is also solved, \ie\ 
the degeneracy of scalar fermion masses, which leads to
the suppression of the flavor changing neutral current (FCNC) processes 
via SUSY particles, can be realized.\cite{horiz} 
In the scenario, $E_6$ GUT
is more interesting than $SO(10)$ GUT, because $E_6$ models suppress 
more naturally the FCNC processes.
Moreover, in $E_6\times SU(3)_H$ GUT, all the three generation 
quarks and leptons can be unified into a single multiplet 
$({\bf 27},{\bf 3})$, while keeping the above attractive points,
including bi-large neutrino mixing angles.

In the analysis, it is important that all the scales of the vacuum 
expectation values~(VEVs) are determined by the anomalous $U(1)_A$ 
charges.
To be more precise, VEVs of GUT gauge singlet operators 
~($G$-singlets) $O_i$ with anomalous $U(1)_A$ charges $o_i$ are 
generally given by
\begin{equation}
\VEV{O_i}\sim \left\{ 
\begin{array}{ccl}
  \lambda^{-o_i} & \quad & o_i\leq 0 \\
  0              & \quad & o_i>0
\end{array} \right. , 
\label{VEV}
\end{equation}
if they are determined by $F$-flatness conditions.
Here $\lambda(\ll 1)$ is the ratio of the cutoff scale $\Lambda$ and 
the VEV of the FN field $\Theta$, whose anomalous $U(1)_A$ charge 
is normalized to $-1$. 
Throughout this paper, we use units in which $\Lambda=1$, and 
denote all the superfields and chiral operators by uppercase 
letters and their anomalous $U(1)_A$ charges by the corresponding 
lowercase letters.
This vacuum structure (\ref{VEV}) is essential for 
the natural gauge coupling unification.\cite{NGCU}
The reason why VEVs are determined as (\ref{VEV}) is 
explained in detail in Ref.\citen{E6,NGCU,maekawa2,horiz,BM,maekawa}, 
and here, we only figure out the discussion using the simplest case, 
in which all the fields, $Z_i$, are gauge singlet. 
Roughly speaking, theories with anomalous $U(1)_A$ gauge symmetry with
Fayet-Iliopoulos $D$-term have two kinds of vacua; 
(1) $\VEV{Z}\sim \order1$ and (2) $\VEV{Z_i^+}=0 (z_i^+>0)$, 
if all the terms allowed by the symmetry are introduced 
in the superpotential $W\Ll Z_i\Rl$ with $\order1$ coefficients.
In the second vacua, if the coefficient of Fayet-Iliopoulos $D$-term 
$\xi$ is smaller than 1, all the VEVs of negatively charged fields 
$Z_i^-$ must be smaller than 1, because $D$-flatness condition of 
the anomalous $U(1)_A$ gauge symmetry becomes
\begin{equation}
 D_A=g_A \left(\sum_i z^-_i |Z^-_i|^2 +\xi^2 \right)=0.
\label{eq:dflat}
\end{equation}
In the followings, we concentrate on the second vacua,
in which the hierarchical structure of Yukawa couplings 
can be understood by the FN mechanism. 
Note that in the vacua, $F$-flatness conditions of 
negatively charged fields $F_{Z_i^-}=0$ become trivial.
And in the $F$-flatness conditions of positively charged fields 
$F_{Z^+}=0$, only a part of superpotential, which are linear 
in positively charged fields, is required in determining the VEVs of
negatively charged fields.
This is because the interactions which include more than two 
positively charged fields give, in the $F$-flatness conditions,
the terms which include at least 
one vanishing VEVs of positively charged field 
and therefore become irrelevant.
Note that in the superpotential, the coefficient for an interaction $O$
is determined by its anomalous $U(1)_A$ charge $o$ as $\lambda^o$.
Hence, if we rewrite everything in terms of 
neutral operators, $\hat Z_i \equiv \Theta^{z_i}Z_i$, under anomalous $U(1)_A$ 
symmetry,  all the coefficients become $\order1$.
And VEVs of $\hat Z_i$ are generally expected to be $\order1$, 
which lead $\VEV{ Z_i} \sim \lambda ^{-z_i}$.
As for non-singlet fields, the $D$-flatness conditions of GUT symmetry 
further determine VEVs of fields. 
For example, the VEVs of a field $C$ in a complex representation and of its 
mirror field $\bar C$ can be obtained from  the VEV of the $G$-singlet 
$\VEV{\bar CC}\sim \lambda^{-(c+\bar c)}$ and $D$-flatness condition
of GUT symmetry as 
$\left|\VEV{C}\right|=\left|\VEV{\bar C}\right|
         \sim \lambda^{-(c+\bar c)/2}$.
Here, we would stress that the non-singlet field $C$ may acquire a
non-vanishing VEV even when $c>0$, if $c+\bar c<0$.

Unfortunately, in the previously proposed model,\cite{E6} the gauge 
coupling at the GUT scale tend to be in non-perturbative region. 
This is because the Higgs sector include many Higgs fields. 
Although there is a possibility that they remain in perturbative 
region due to the ambiguities of $\order1$ coefficients and/or the
freedom of the charge assignment, 
it is important to search other $E_6$ Higgs sector
which has simpler Higgs contents. 
In this paper, we examine simpler \E6 models and construct 
models consistent with the already proposed matter sector.\cite{BM}
In these models, the sliding singlet mechanism,\cite{SS} 
as well as the Dimopoulos-Wilczek~(DW) mechanism,\cite{DW} 
acts to realize naturally DT splitting.

\section{Overview of $SO(10)$ and $E_6$ GUT} \label{overview}
Here we make a quick review of the $SO(10)$  unified scenario%
,\cite{maekawa} and the $E_6$ unified model proposed
previously.\cite{E6,BM}

\subsection{$SO(10)$ Higgs sector}
The content of the Higgs sector in $SO(10)\times U(1)_A$ is 
listed in Table I. 
\begin{table}
\begin{center}
Table I. Typical values of anomalous $U(1)_A$ charges.\\
\vspace{1mm}
\begin{tabular}{|c|c|c|} 
\hline
                  &   non-vanishing VEV  & vanishing VEV \\
\hline 
{\bf 45}          &   $A(a=-1,-)$        & $A'(a'=3,-)$      \\
{\bf 16}          &   $C(c=-3,+)$        
                  & $C'(c'=2,-)$      \\
${\bf \overline{16}}$&$\bar C(\bar c=0,+)$ 
                  & $\bar C'(\bar c'=5,-)$ \\
{\bf 10}          &   $H(h=-3,+)$        & $H'(h'=4,-)$      \\
{\bf 1}           &$\Theta(\theta=-1,+)$,$Z_i(z_i=-2,-)$\ $(i=1,2)$
                  & $S(s=3,+)$ \\
\hline
\end{tabular}
\end{center}
\end{table}
Here the symbols $\pm$ denote the $Z_2$ parity.
Following the general discussion of the VEVs,
positively charged Higgs $A'$, $C'$, $\bar C'$, $H'$ and
$S$ have vanishing VEVs. The superpotential required in 
determining of the VEVs can be written
\begin{equation}
W=W_{H^\prime}+ W_{A^\prime} + W_S + W_{C^\prime}+W_{\bar C^\prime}.
\end{equation}
Here $W_X$ denotes the terms linear in the positively charged field
$X$, which has vanishing VEV.  Note that terms 
including two fields with vanishing VEVs like 
$\lambda^{2h^\prime}H^\prime H^\prime$
give contributions to the mass terms but not to the $F$-flatness conditions
to determine the VEVs.
Examining 
\begin{equation}
W_{A'}=\tr A'A+\tr A'A\tr A^2+\tr A'A^3,
\label{WAso10}
\footnote{
We often omit the $\order1$ coefficient and the power of $\lambda$ 
, which is easily understood from anomalous $U(1)_A$ charge 
{\it e.g.} $\lambda^{a'+a}$ for $A'A$.
}
\end{equation} 
the adjoint Higgs
field $A$ can have the VEV 
$\VEV{A({\bf 45})}_{B-L}=\tau_2\times {\rm diag}
(v,v,v,0,0)$, which breaks $SO(10)$ into
$SU(3)_C\times SU(2)_L\times SU(2)_R\times U(1)_{B-L}$.
Here, note that if the third term of r.h.s. of (\ref{WAso10}) 
are absent, the corresponding $F$-flatness condition determines only 
$\tr A^2$, 
and we cannot expect to obtain naturally this DW form of the VEV.
This form of the VEV plays an
important role in solving the DT splitting problem through
the interaction
\begin{equation}
W=H'AH,
\label{WHso10}
\end{equation}
that gives non-vanishing mass term only for triplet Higgs, but not for
doublet Higgs.\cite{DW} 
Because a mass term $H'^2$ is allowed by the symmetry,
only one pair of doublet Higgs in $H$ field becomes massless, \ie\ 
the DT splitting is realized.
The spinor Higgs fields 
$C$ and $\bar C$ must have non-vanishing VEVs because of
the $F$-flatness condition of the superpotential
\begin{equation}
W_{S}=S(1+\bar CC+\tr A^2).
\label{WSso10}
\end{equation} 
The $F$-flatness conditions of the superpotential
\begin{eqnarray}
W_{C'}&=&\bar C(A+Z_i)C',\label{WCso10}\\
W_{\bar C'}&=& \bar C'(A+ Z_i)C
\label{WbCso10}
\end{eqnarray}
can align the VEVs $\VEV{C}=\VEV{\bar C}
(=\lambda^{-(c+\bar c)/2})$, which break
$SU(2)_R\times U(1)_{B-L}$ into $U(1)_Y$, 
with the VEV $\VEV A$.\cite{BarrRaby}

It is obvious that the mass term $H'H$ spoils the DT splitting.
Therefore, this mass term must be forbidden. Unfortunately,
this term cannot forbidden by the SUSY zero mechanism, because
the important term $H'AH$, whose charge is smaller than the 
charge of $H'H$, must be allowed by the symmetry. Therefore,
we introduce $Z_2$ symmetry to forbid the term $H'H$.

There are several terms that must be forbidden for the stability 
of the DW mechanism. For example, $H^2$ and 
$H Z H^\prime$ induce a large mass of the doublet Higgs, 
and the term $\bar CA^\prime A C$ would destabilize the DW form of 
$\VEV{A}$.
We can easily forbid these terms using the SUSY zero mechanism.
For example, if we choose
$h<0$, then $H^2$ is forbidden, and if we choose 
$\bar c+c+a+a^\prime<0$, then
$\bar CA^\prime A C$ is forbidden. 
Once these dangerous terms are forbidden
by the SUSY zero mechanism, higher-dimensional terms that 
also become dangerous (for example, 
$\bar CA^\prime A^3 C$ and $\bar CA^\prime C\bar CA C$) are 
automatically forbidden, because only negatively charged $G$-singlets have
non-vanishing VEVs.
This is also an advantage of our scenario. 

For the quark and lepton sector, we introduced four superfields 
$\Psi_i({\bf 16})$ $(i=1,2,3)$ and $T({\bf 10})$ with typical values of 
the charges $(\psi_1,\psi_2,\psi_3)=(9/2,7/2,3/2)$ and $t=5/2$.
The half integer charges for matter fields play the same role 
as R-parity.
Because the $SO(10)$ representations ${\bf 16}$ and ${\bf 10}$ are 
divided as
\beqn
  {\bf 16} &\rightarrow& {\bf 10}_1+{\bf \bar 5}_{-3}+{\bf 1}_5 \\
  {\bf 10} &\rightarrow& {\bf 5}_{-2}+{\bf \bar 5}_2 
\eeqn
under $SU(5)\times U(1)_V(\subset SO(10))$,
a liner combination of four {$\bf\bar5$} of $SU(5)$ becomes massive
with the {\bf 5} component in $T$.
Therefore, three massless modes of ${\bf \bar 5}$ are linear 
combinations of four ${\bf \bar 5}$. 
Under the typical charge assignment, three massless modes become 
$({\bf \bar 5}_{\Psi_1},
{\bf \bar 5}_T+\lambda^\Delta {\bf \bar 5}_{\Psi_3}, 
{\bf \bar 5}_{\Psi_2})$.
Here because the field ${\bf \bar 5}_T$ has no Yukawa couplings with the 
Higgs $H$, we wrote the mixing with ${\bf \bar 5}_{\Psi_3}$, through which
the massless mode ${\bf \bar 5}_T+\lambda^\Delta {\bf \bar 5}_{\Psi_3}$
can have Yukawa couplings. The mixing parameter $\Delta$ is obtained by 
$\Delta\equiv t-\psi_3+\frac{1}{2}(c-\bar c)=\frac{5}{2}$.
The neutrino majorana masses are given by $\Psi_i\Psi_j\bar C\bar C$, 
and the mass of the heaviest light neutrino is written 
\bequ
  m_{\nu_3} \sim \lambda^{-(l+5)} \frac{\VEV{H_u}^2\eta^2}
                                   {\Lambda}, 
\eequ
where $\eta$ is a renormalization factor and 
$l=-(h+c-\bar c+9)=-3$. 
For $\lambda\sim0.22$ and $\VEV{H_u}\eta=100$-$200\GeV$, 
$-1<l<-4$ is needed for correct atmospheric neutrino mass scale, 
\cite{Matm} 
though the requirement depends on the ambiguity of $\order1$ 
coefficients.

\subsection{$E_6$ Higgs sector}

In this subsection, we recall the $E_6$ Higgs sector proposed 
previously.\cite{E6}
The content of the Higgs sector with $E_6\times U(1)_A$ gauge 
symmetry is given in Table \II, where the symbols $\pm$ denote the 
$Z_2$ parity quantum numbers.%
\footnote{
  Here the composite operator $\bar\Phi\Phi$ play the role of 
 the FN field $\Theta$.
} 
\begin{table}
\begin{center}
Table \II. Typical values of anomalous $U(1)_A$ charges.
\\ \vspace{0.1cm}
\begin{tabular}{|c|c|c|} 
\hline
                  &   non-vanishing VEV  & vanishing VEV \\
\hline 
{\bf 78}          &   $A(a=-1,-)$        & $A'(a'=4,-)$      \\
{\bf 27}          &   $\Phi(\phi=-3,+)$\  $C(c=-6,+)$ &  $C'(c'=7,-)$  \\
${\bf \overline{27}}$ & $\bar \Phi(\bar \phi=2,+)$ \  $\bar C(\bar c=-2,+)$ &

                  $\bar C'(\bar c'=8,-)$ \\
{\bf 1}           &   $Z_i(z_i=-2,-)$\ $(i=1,2,3)$ \   &  \\
\hline
\end{tabular}
\end{center}
\end{table}
To explain how to embed the previous $SO(10)$ model into 
the $E_6$ model, it is helpful to see that the adjoint 
representation ${\bf 78}$ and the fundamental representation 
${\bf 27}$ are divided as
\begin{eqnarray}
{\bf 78}&\rightarrow& {\bf 45}_0+{\bf 16}_{-3}+{\bf \overline{16}}_3
        +{\bf 1}_0, \\
{\bf 27}&\rightarrow& {\bf 16}_1+{\bf 10}_{-2}+{\bf 1}_4
\end{eqnarray}
under $SO(10)\times U(1)_{V'}(\subset E_6)$.
The non-vanishing VEVs $|\VEV{\Phi}|=|\VEV{\bar \Phi}|$ break
$E_6$ into $SO(10)$. The VEVs $\VEV{A}$ and 
$|\VEV{C}|=|\VEV{\bar C}|$ break the $SO(10)$ into the 
standard model gauge group, as in the previous 
$SO(10)$ GUT.
Note that Higgs sector has the same number of superfields 
in non-trivial representation as the $SO(10)$ Higgs sector, 
in spite of the fact that the larger group $E_6$ requires 
additional Higgs fields to break $E_6$ into the $SO(10)$ gauge group.
Therefore, in a sense, this
$E_6$ Higgs sector unifies the $SO(10)$ Higgs sector.
Actually, the Higgs fields $H$ and $H'$ of the $SO(10)$ model 
are contained in $\Phi$ and $C'$, respectively, 
in this $E_6$ model. 

Here non-vanishing VEVs are determined by 
\begin{equation}
  W=W_{A^\prime} + W_{C^\prime}+W_{\bar C^\prime}.
\end{equation}
We, however, have to be careful about dealing with $W_{A^\prime}$. 
Due to a characteristic of the \E6 group, 
$W_{A^\prime}=A'(A+A^3)$ does not contain $\tr{\bf 45}_A'{\bf 45}_A^3$ 
and the DW form cannot obtained in a natural way.
(We call this ``factorization problem''.)
Because this is caused from \E6 characteristic, if an \E6 breaking 
effect couples to $A'A^3$, \eg~$\bar\Phi A'A^3\Phi$, 
this problem is avoided.
$W_{C^\prime}$ and $W_{\bar C^\prime}$ play a similar role as denoted
in (\ref{WHso10})-(\ref{WbCso10}).

For the quark and lepton sector,\cite{BM} we introduced 
three superfields $\Psi_i({\bf 27})$ $(i=1,2,3)$ (with typical values of 
the charges $(\psi_1,\psi_2,\psi_3)=(9/2,7/2,3/2)$), in which all 
the superfields in quark and lepton sector of the $SO(10)$ model 
are embedded.
And this minimal content can realizes realistic 
quark and lepton mass matrices including bi-large 
neutrino mixings in a similar manner.
Here we define a parameter $r$ as 
\begin{equation}
\lambda^r\equiv \lambda^{c}\VEV{C}/\lambda^\phi\VEV{\Phi}
=\lambda^{\frac{1}{2}(\bar c-c+\phi-\bar\phi)}.
\end{equation}
This parameterizes the mixing of ${\bf\bar5}$ matter as $\Delta=3-r$ 
and must be around $0<r<3/2$ for bi-large neutrino mixings. 
This requirement is also depends on the ambiguity of 
$\order1$ coefficients.

The charge assignment in Table \II\ provides a complete $E_6$
GUT with $l=-2$ and $r=1/2$, 
although the gauge coupling at the cutoff scale 
may be in non-perturbative region.\cite{NGCU}

\section{simpler $E_6$ Higgs sectors}

In the previous \E6 model, $C({\bf 16})$ and $\bar C({\bf\cc{16}})$ of 
$SO(10)$ model are embedded into ${\bf 27}$ field and 
${\bf \overline{27}}$ field, respectively. 
However, they may also be embedded into ${\bf 78}$ field, 
resulting simpler $E_6$ models. 
Here we examine this alternative embedding.

Since we introduce two adjoint Higgs $A'$ and $A$, we have two kinds of 
possibilities for reducing the Higgs sector.
\begin{enumerate}
\item
  The VEV $\VEV{{\bf 16}_{A'}}$ or $\VEV{{\bf \overline{16}}_{A'}}$ 
  is non-vanishing.
\item
  The VEV $\VEV{{\bf 16}_{A}}$ or $\VEV{{\bf \overline{16}}_{A}}$ 
  is non-vanishing.
\end{enumerate}
Note that it must be forbidden that ${\bf 16}$ and 
${\bf \overline{16}}$ have non-vanishing VEVs simultaneously, 
which destabilizes the DW form of VEVs. 
For example, if the VEVs $\VEV{\bf 16_{A'}}$ and 
$\VEV{\bf \overline{16}_{A'}}$ are non-vanishing, 
the interactions $A'^n$ destabilize the DW form of VEVs 
because $F_{{\bf 45}_{A'}}$ includes the VEVs $\VEV{\bf 16_{A'}}$ and 
$\VEV{\bf \overline{16}_{A'}}$. 
At first glance, such an asymmetric VEV structure is forbidden 
by $D$-flatness conditions.
But it is shown below that such an interesting VEV can satisfy
the $D$-flatness conditions.

\subsection{$\VEV{{\bf 16}_{A'}}\neq 0$}

The typical Higgs content is represented in Table \III.

\begin{table}
\begin{center}
Table \III. Typical values of anomalous $U(1)_A$ charges.
\\ \vspace{0.1cm}
\begin{tabular}{|c|c|} 
\hline 
{\bf 78}          &   $A(a=-1)$\ $A'(a'=5)$       \\
{\bf 27}          &   $\Phi(\phi=-5)$\  $C'(c'=7)$  \\
${\bf \overline{27}}$ &$\bar \Phi'(\bar \phi'=6)$ \ $\bar C(\bar c=-6)$  \\
{\bf 1}           &   $\Theta(\theta=-1)$\ $Z_i(z_i=-1)$ \ ($i=$1-5)\ 
                      $S(s=6)$   \\
\hline
\end{tabular}
\end{center}
\end{table}
Suppose that among the above Higgs fields, only ${\bf 45}_A$, 
${\bf 1}_\Phi$, ${\bf \overline{16}}_{\bar C}$ and
${\bf 16}_{A'}$ have non-vanishing VEVs such as
\begin{eqnarray}
  \VEV{{\bf 45}_A}&=&\tau_2\times{\rm diag} (v,v,v,0,0)
    \quad (v\sim \lambda^{-a}),\\
  |\VEV{{\bf 1}_\Phi}|&=&|\VEV{\bf \overline{16}}_{\bar C}|
                       =|\VEV{\bf 16}_{A'}|
                       \sim \lambda^{-\frac{1}{3}(\bar c+a'+\phi)}.
\label{VEVap}
\end{eqnarray}
As mentioned above, if $\phi+a'+\bar c<0$, the VEV $\VEV{\bar CA'\Phi}$
can be non-vanishing, which means $A'$ has a non-vanishing VEV. 
Actually, this vacuum satisfies the relations
$\VEV{\tr A'^n}=0$ and 
$\VEV{\bar CA'\Phi}\sim\lambda^{-(\bar c+a'+\phi)}$, which are 
consistent with the VEV relation (\ref{VEV}).
And this vacuum satisfies not only the $D$-flatness conditions
for $SO(10)$ but also that of $U(1)_{V'}$ 
\begin{equation}
D_{V'}\ :\  4|{\bf 1}_{\Phi}|^2-3|{\bf 16}_{A'}|^2
               -|{\bf \overline{16}}_{\bar C}|^2=0.
\end{equation}
Therefore, it is obvious that this vacuum satisfies all the $E_6$
$D$-flatness conditions.

Next we discuss the $F$-flatness conditions to know how such 
a vacuum can be obtained. 
For simplicity, we assume that any component fields other than 
${\bf 45}_A$, ${\bf 1}_\Phi$, ${\bf \overline{16}}_{\bar C}$ and
${\bf 16}_{A'}$ have vanishing VEVs. To determine the VEV of ${\bf 45}_A$,
the superpotential
\begin{equation}
W_{A'}=A'A+A'A^3+A'A^4+A'A^5
\end{equation}
is sufficient to be considered. 
Here, for simplicity, singlets $Z_i$ are not written explicitly.
The $F$-flatness condition of 
${\bf 45}_{A'}$ leads to the DW type of VEV,
$\VEV{{\bf 45}_A}\sim \tau_2\times{\rm diag} (v,v,v,0,0)$. 
(Here $A'A^5$ is needed to avoid the ``factorization problem''.)
Because the positively charged field $A'$ has a non-vanishing VEV 
$\VEV{{\bf 16}_{A'}}\neq 0$, the $F$-flatness conditions of the
negatively charged fields may become non-trivial conditions. 
Fortunately, in this model, there is no such a non-trivial condition,
for example, $F_{{\bf \overline{16}}_A}=0$ is trivial because 
${\bf \overline{16}}_A$ is a Nambu-Goldstone mode in the superpotential
$W_{A'}$.
The $F$-flatness condition of $S$, which is obtained from the 
superpotential
\begin{equation}
W_S=S(1+\bar CA'\Phi+f_S(A,Z_i)),
\end{equation}
leads to
\begin{equation}
\VEV{\bar CA'\Phi}\sim \lambda^{-(\bar c+a'+\phi)}.
\end{equation}
The $D$-flatness conditions of $SO(10)$ and $U(1)_{V'}$ lead to
\begin{equation}
  |\VEV{{\bf 1}_\Phi}|=|\VEV{{\bf \overline{16}}_{\bar C}}|
                      =|\VEV{{\bf 16}_{A'}}|
                    \sim \lambda^{-\frac{1}{3}(\bar c+a'+\phi)},
\end{equation}
which are the desired vacuum in Eq. (\ref{VEVap}).

The $F$-flatness conditions of $C'$, which are obtained from the
superpotential
\begin{equation}
W_{C'}=\bar C(1+Z_i+A+A'(f_C(A, Z_i)+\bar CA'\Phi)C',
\end{equation}
are written
\begin{eqnarray}
F_{{\bf 16}_{C'}}&=&(1+Z_i+A){\bf\cc{16}}_{\bar C}=0, \\
F_{{\bf 1}_{C'}}&=&
(f_C(A, Z_i)+\bar CA'\Phi){\bf\cc{16}}_{\bar C}{\bf16}_{A'}=0.
\label{alignment}
\end{eqnarray}
These conditions realize an alignment between the VEVs 
$\VEV{{\bf 45}_A}$, $\VEV{{\bf \overline{16}}_{\bar C}}$ and 
$\VEV{{\bf 16}_{A'}}$ by shifting the VEVs of singlet fields $Z_i$ 
and as the result, the pseudo Nambu-Goldstone fields become massive.
The $F$-flatness condition of ${\bf \overline{16}}_{A'}$, which is obtained 
from the superpotential
\begin{equation}
W_{A'A'}=A'(f_A(A,Z_i)+\bar CA'\Phi)A',
\end{equation}
realizes also an alignment between $\VEV{{\bf 45}_A}$
 and $\VEV{{\bf 16}_{A'}}$.

It is interesting that in this model, the sliding singlet mechanism%
\cite{SS} is naturally realized.
The $F$-flatness conditions of $\bar \Phi'$, which are obtained from the
superpotential
\begin{equation}
W_{\bar\Phi'}=\bar \Phi'(1+Z_i+A+A'(f_\Phi(A, Z_i)+\bar CA'\Phi))\Phi,
\end{equation}
are written
\begin{eqnarray}
F_{{\bf 1}_{\bar \Phi'}}&=&(1+Z_i){\bf1}_\Phi=0, \\
F_{{\bf \overline{16}}_{\bar \Phi'}}&=&
(f_\Phi(A, Z_i)+\bar CA'\Phi){\bf1}_\Phi {\bf16}_{A'}=0.
\label{sliding}
\end{eqnarray}
At first glance, the component field ${\bf 10}_{\Phi}$, 
which includes doublet Higgs in the standard model, 
seems to have a mass term from the superpotential 
$\bar \Phi'(1+Z_i)\Phi$. 
However, this mass term vanishes in the desired vacuum 
which satisfy the first condition in the above $F$-flatness conditions. 
Moreover, because the adjoint field $A$ has the DW type of VEV, 
only the triplet Higgs becomes massive through the interaction
$\bar \Phi'A\Phi$. As the result, DT splitting is realized
by the sliding singlet mechanism and the DW mechanism.
The essential point here is that the doublet Higgs has the same 
quantum number under the generator $\VEV{A}$ as the component field 
${\bf 1}_\Phi$ which has non-vanishing VEV. 
This mechanism can be generalized, that will be discussed
in the following paper.\cite{SSE}

In the above model, for intelligibility, we introduced a positively 
charged singlet $S$ in order to fix the VEV 
$\VEV{\bar CA'\Phi}\sim \lambda^{-(\bar c+a'+\phi)}$. However, 
one of the non-trivial $F$-flatness conditions of ${\bf 1}_{C'}$,
${\bf \overline{16}}_{A'}$ and ${\bf \overline{16}}_{\bar \Phi'}$
can play the same role as $S$. If we do not introduce the field $S$,
the number of the negatively charged singlets becomes four.

It is worthwhile to note how to determine the anomalous $U(1)_A$ 
charges.
In order to realize DT splitting, the terms
\begin{equation}
A'A^5,\bar \Phi'A\Phi, \bar C(A+Z)C'
\end{equation}
must be allowed, and the term
\begin{equation}
\bar CA'^2\Phi
\end{equation}
must be forbidden. 
These requirements can be rewritten as the inequalities.
We determined the charges in order to satisfy the inequalities.

Unfortunately, we have not found realistic matter sector with this
Higgs sector. Actually, the mixing parameter $r$, which is 
obtained by
\begin{equation}
\lambda^r\equiv\frac{\lambda^{a'+\phi}\VEV{{\bf16}_{A'}{\bf1}_\Phi}}
                    {\lambda^{\phi}\VEV{{\bf1}_\Phi}}
               =\lambda^{a'}\VEV{{\bf16}_{A'}}
               =\lambda^{\frac{1}{3}(2a'-\bar c-\phi)},
\end{equation}
must be around 1/2 in order to obtain bi-large neutrino mixings, and
it looks to be impossible, because $2a'-\bar c-\phi\gg 1$.

\subsection{$\VEV{{\bf16}_A}\neq0$}

In this section, we consider another possibilities in which 
$C({\bf16})$ of the $SO(10)$ model is embedded into negatively 
charged adjoint Higgs $A({\bf78})$. This possibility is more promising
because the condition for realistic matter sector, 
$2a-\bar c-\phi\sim 1$, can be realized.
The content of the Higgs sector is the same as 
in the previous possibility, except for the charges and 
the number of singlets.

To begin with, we examine $D$-flatness conditions.
Because $\VEV{{\bf45}_A}\neq0$ and 
$\VEV{{\bf16}_A}\neq0$, 
the $D$-flatness condition of ${\bf\cc{16}}$ 
direction gives a non-trivial condition.
In order to compensate the contribution from $A$ in the
condition, $\Phi$ and/or 
$\bar C$ must have non-zero VEV in both {\bf1} and {\bf16}
(${\bf\cc{16}}$) components.
Therefore, non-trivial $D$-flatness conditions are 
\beqn
  D_{V+V'} &\quad:\quad& \abs{{{\bf1}_\Phi}}^2  =  
      \abs{{{\bf16}_A}}^2 + \abs{{{\bf1}_{\bar C}}}^2, 
\label{DVVp} \\
  D_V &\quad:\quad& \abs{{{\bf\cc{16}}_{\bar C}}}^2 = 
      \abs{{{\bf16}_A}}^2 + \abs{{{\bf16}_\Phi}}^2, 
\label{DV} \\
  D_{\cc{\bf16}} &\quad:\quad& {{\bf45}_A}^*{{\bf16}_A}
        = {{\bf1}_\Phi}^*{{\bf16}_\Phi}
         -{{\bf\cc{16}}_{\bar C}}^*{{\bf1}_{\bar C}}. 
\label{D16} 
\eeqn
In addition, we suppose 
\beqn
  \VEV{{\bf\cc{16}}_{\bar C}}\VEV{{\bf16}_A}\VEV{{\bf1}_\Phi} \nn
  &\sim& 
  \VEV{{\bf\cc{16}}_{\bar C}}\VEV{{\bf45}_A}\VEV{{\bf16}_\Phi} \nn\\
  &\sim& 
  \VEV{{\bf{1}}_{\bar C}}\VEV{{\bf45}_A}\VEV{{\bf1}_\Phi} \nn\\
  &\sim& 
  \lambda^{-(\bar c+a+\phi)}
  \,\,\equiv\,\, \lambda^{-3k} 
\label{3Fcondi}\\
  \VEV{{\bf45}_A} &\sim& \lambda^{-a} 
\label{AFcondi}
\eeqn
are obtained from $F$-flatness conditions as generally expected.%
\footnote{
Strictly speaking, if three conditions in Eq. (\ref{3Fcondi}) were 
determined by $F$-flatness conditions, the $F$-flatness and $D$-flatness 
conditions would become over-determined. 
Therefore, only two of the three conditions are
determined by $F$-flatness conditions. Then, another solution,
\begin{equation}
\VEV{{\bf 1}_A}\sim \VEV{{\bf 16}_A}\sim \lambda^{-a}\ll
\VEV{{\bf 1}_\Phi}\sim \VEV{{\bf 16}_\Phi}\sim \VEV{{\bf 1}_{\bar C}}
\sim \VEV{{\bf 16}_{\bar C}},
\nn
\end{equation}
may appear, by which the natural gauge coupling unification is 
not realized.
Though the $\order1$ coefficients determine which vacuum is realized,
the desired vacuum is obtained in some (finite) region of parameter space
of the $\order1$ coefficients.
}
From these conditions except for Eq.(\ref{D16}), 
the orders of VEVs are determined 
as follows, for $\lambda^{-a}\gg\lambda^{-k}$.
\beqn
  &\VEV{{\bf\cc{16}}_{\bar C}} 
    \sim \VEV{{\bf16}_A}
    \sim \VEV{{\bf1}_\Phi}
    \sim \lambda^{-k} 
    \equiv \lambda^{-a}\lambda^r & \label{VEV1}\\
  &\VEV{{\bf{1}}_{\bar C}}
    \sim \VEV{{\bf16}_\Phi}
    \sim \lambda^{a-2k} 
    \sim \lambda^{-a}\lambda^{2r} & \label{VEV2}\\
  &\VEV{{\bf45}_A} \sim \lambda^{-a}& \label{VEV3}
\eeqn
Here, $r=a-k$ is the mixing parameter, introduced in \S\ref{overview}.
For these VEVs, the effective charges can be defined and therefore 
the natural gauge coupling unification is realized.\cite{NGCU}
Taking account of Eq.(\ref{D16}), it may appear that this condition 
requires $r>0$.
However, since $r$ should be small ($\sim1/2$) as mentioned above 
and there is an ambiguity due to order one coefficients,
Eq.(\ref{D16}) can be satisfied.
To be more precise, Eq.(\ref{D16}) has the form 
$\lambda^{-2a+r} = \lambda^{-2a+3r}+\lambda^{-2a+3r}$, 
and the r.h.s may become 
$2\lambda^{-2a+3r}\sim\lambda^{-2a+r}\lambda^{2r-1/2}$, 
allowing $r=1/4$.
And the ambiguities of $\order1$ coefficients makes 
a larger $r$ possible.

Next, we examine $F$-flatness conditions.
The typical charge assignment of Higgs sector is represented 
in Table \IV.
\begin{table}
\begin{center}
Table \IV. Typical values of anomalous $U(1)_A$ charges.
\\ \vspace{0.1cm}
\begin{tabular}{|c|c|} 
\hline 
{\bf 78}          &   $A(a=-1,+)$\ $A'(a'=5,+)$       \\
{\bf 27}          &   $\Phi(\phi=-3,+)$\  $C'(c'=6,-)$  \\
${\bf \overline{27}}$ &$\bar \Phi'(\bar \phi'=5,+)$ \ $\bar C(\bar c=0,-)$  \\
{\bf 1}           &   $\Theta(\theta=-1,+)$\ $Z_i(z_i=-1,+)$\ $(i=1,2)$   \\
\hline
\end{tabular}
\end{center}
\end{table}
Here the VEVs are again determined by 
\begin{equation}
  W=W_{A^\prime} + W_{\bar\Phi^\prime}+W_{C^\prime},
\end{equation}
where 
\beqn
  W_{A'} &=& A'(A+A^3+A^4+A^5) \\
  W_{\bar\Phi'} &=& \bar\Phi'(1+A+Z_i+A^2+AZ_i+Z_i^2)\Phi 
\label{Wphi}\\
  W_{C'} &=& \bar C(1+A+Z_i+\cdots+(\bar C\Phi)^2)C'. 
\eeqn
As in the previous model, the $F$-flatness condition of 
${\bf 45}_{A'}$ leads to the DW type of VEV,
$\VEV{{\bf 45}_A}\sim \tau_2\times{\rm diag} (v,v,v,0,0)$. 
The $F$-flatness condition of ${\bf1}_{\bar\Phi'}$ makes the \E6 singlet 
part in the parenthesis of Eq.(\ref{Wphi}) vanish, leading vanishing 
doublet mass terms (the sliding singlet mechanism).
The $F$-flatness condition of ${\bf\cc{16}}_{\bar\Phi'}$ gives a factored 
equation
\bequ
  (1+A+Z_i)\Ll {\bf45}_A{\bf16}_\Phi + {\bf16}_A{\bf1}_\Phi\Rl
    = 0, 
\eequ
which can be checked by the explicit calculation based on 
\E6 group theory. 
The above two $F$-flatness conditions are satisfied by shifting
the VEVs of two singlets $Z_i$. 
The two $F$-flatness conditions of ${\bf 1}_{C'}$ and ${\bf 16}_{C'}$
and the three $D$-flatness conditions in Eqs. (\ref{DVVp})-(\ref{D16}) 
determine the five VEVs ${\bf 16}_A$, ${\bf 1}_{\Phi}$,
${\bf 16}_\Phi$, ${\bf 1}_{\bar C}$ and ${\bf \overline{16}}_{\bar C}$.
It is straightforward to analyse the mass matrices of Higgs 
to check all modes are superheavy except for one doublet Higgs pair 
contained in ${\bf10}_\Phi$.\footnote{
We would emphasize in this model all the singlet fields also 
become superheavy, while
in the previous models, 
one massless singlet field appears.}

Now, we examine the condition to be compatible with the matter 
sector, for which we introduced the same three superfields 
as in the \E6 models in \S\ref{overview}. 
Applying the discussion to this case, the 
parameters $r$ and $l$ are given from following relations;
\beqn
  &\lambda^r \sim \frac{\lambda^c \VEV{{\bf{16}}_C}}
                       {\lambda^\phi \VEV{{\bf1}_\Phi}}
             \sim \frac{\lambda^{a+\phi} \VEV{{\bf{16}}_A}
                                         \VEV{{\bf1}_\Phi}}
                       {\lambda^\phi\VEV{{\bf1}_\Phi}}
             = \lambda^{a-k}, & \\
  &\lambda^{-(5+l)} \sim \lambda^{4+\phi-2\bar c}
                         \VEV{{\bf\cc{16}}_{\bar C}}^{-2}
                    \sim \lambda^{4+\phi-2\bar c +2k}.&
\eeqn
For example, a set of charges $(a,\phi, \bar c)=(-1, -3, 0)$ as in 
Table \IV\ gives 
$(r, l)=(1/3, -10/3)$, which is allowed as shown in \S\ref{overview}. 

It is worthwhile to note how to determine the symmetry 
in this model. For this purpose, we have to mention 
which terms are needed and which must be forbidden. 
In order to stabilize the DW-type VEV, $\bar CA'\Phi$ 
must be forbidden, and to avoid the 
``factorization problem'', $A'A^5$ is needed.
However, it is difficult to forbid $\bar CA'\Phi$ while allowing 
$A'A^5$ by the SUSY-zero mechanism for small $r (=\frac{1}{3}
(\bar c+\phi-2a))$.
Therefore we need another mechanism to forbid $\bar CA'\Phi$, 
{\it e.g.} to introduce an additional $Z_N$ symmetry.
Since $\Psi_3\Psi_3\Phi$ gives order one top Yukawa 
coupling, this three point interaction should be allowed by the 
$Z_N$ symmetry. 
Thus, we assign nontrivial $Z_N$ charges only on $\bar C$ (and $C'$).
In $W_{\bar{\Phi}'}$, the would-be simplest superpotential 
$W_{\bar{\Phi}'}=\bar{\Phi}'A\Phi$ is not consistent with the 
$D$-flatness conditions for these VEVs (\ref{VEV1})-(\ref{VEV3}).
This can be checked from the explicit calculation in terms of 
$SU(3)_R\times U(1)_{T_L^8}$ ($\subset$\E6), which contain $U(1)_{B-L}$, 
(or, more easily, from examining the mass matrix of 
${\bf10}\times{\bf\cc{10}}$ of $SU(5)$).
This is again a characteristic of \E6 group, and therefore, 
\E6 breaking effects such as $\bar{\Phi}'(AZ+A^2)\Phi$ are needed.
And in order that the sliding singlet mechanism acts, 
$\bar{\Phi}'\bar C\Phi\Phi$ must be forbidden.\cite{SSE}
If $\bar C\bar C\Phi C'$ is forbidden, the $F$-flatness conditions 
of ${\bf 1}_{C'}$ and ${\bf 16}_{C'}$ cannot fix the scale of the VEVs
of $\bar C$ and $\Phi$. It means that two other conditions to fix these
VEVs are required. Though introducing two additional positively charged
singlets (and two additional negatively charged singlets) makes it 
possible, it may be more simple to introduce the term $\bar C\bar C\Phi C'$.
If we take the non-trivial $Z_N$ charges for $(\bar C, C')$ as
$(1, N-1)$, the term becomes $\bar C(\bar C\Phi)^NC'$.
To allow the terms $A'A^5$, $\bar \Phi'AZ\Phi$ and 
$\bar C(\bar C\Phi)^NC'$, we adopt 
the anomalous $U(1)$ charges  
$(a', \bar \phi', c')$%
$=(-5a, -\phi-a-z, -\bar c-N(\bar c+\phi))$.
If we take $(a,\phi,\bar c)=(-1,-3,0)$, then  
$(a', \bar \phi', c')=(-1, 5, 5, 3N)$ and 
a consistent model is constructed. 
In this model, $N=2$ is sufficient to decouple $\Phi$ and $\bar C$ 
from $W_{A'}$.
(See Table \IV.) 
As in the \E6 models in \S\ref{overview},  
the half-integer charges of matter supermultiplets play
the same role as ``R-parity'' in this model.
Other charge assignments
$(a,\phi, \bar c, z_i, a', \bar \phi', c')=$ 
$(-1/2, -3, 4/3, -1/2, 5/2, 9/2, 11/3)$ and 
$(-1, -3, 1/3, -1, 5, 5, 23/3)$  give other examples of 
consistent models.
Although the former requires the ``R-parity'' (or $Z_2$ symmetry same 
as in \S\ref{overview}), the latter requires no additional $Z_N$ 
symmetry.

\section{Summary and Discussion} 

In this paper, we consider more simple version of \E6 model with 
anomalous $U(1)$ symmetry than already proposed model.\cite{E6} 
Here, the spinor Higgs $C(\bf{16})$ of $SO(10)$ model 
is embedded into \E6 adjoint Higgs $A'$ or $A$.
Unfortunately, the former case is incompatible with the matter 
sector as far as we know, although the doublet-triplet splitting%
\cite{DTsplitting} and the natural gauge coupling unification%
\cite{NGCU} are realized. 
On the other hand, the latter case is compatible with the matter 
sector, resulting consistent \E6 models. 
Moreover, we can construct a model in which no additional 
symmetry other than the anomalous $U(1)$ symmetry.
This is due to the sliding singlet mechanism,\cite{SS} which we will 
discuss in more detail in another paper.\cite{SSE}
It is interesting that in \E6 model,  both the Dimopoulos-Wilczek 
mechanism\cite{DW} and the sliding singlet mechanism can act elegantly 
to solve the doublet-triplet splitting problem. 
Because we introduced generic interactions, the theory can be defined 
essentially by their symmetry. In the models proposed in this paper,
we introduced only six non-trivial representation fields for the 
Higgs sector and three fields for matter sector. It is surprising that
complete $E_6$ grand unified theories can be constructed by assigning 
only nine (integer) charges.

Due to the smaller Higgs sector, the gauge coupling at 
the cutoff scale tend to remain perturbative region. 
And the evaluation of the lifetime of a nucleon in this model 
$\tau_p(p\rightarrow e^+\pi^0)\sim 5\times 10^{33}$ years\cite{NGCU} 
(for $a=-1$) becomes more reliable and gives strong motivation to search 
the proton decay $p\rightarrow e^+\pi^0$. 

\section*{Acknowledgement}
N.M. thanks M.Bando for valuable discussion, 
and is supported in part by Grants-in-Aid for Scientific 
Research from the Ministry of Education, Culture, Sports, Science 
and Technology of Japan.

\end{document}